\documentclass[conference]{IEEEtran}

\makeatletter % changes the catcode of @ to 11
\newcommand{\linebreakand}{%
  \end{@IEEEauthorhalign}
  \hfill\mbox{}\par
  \mbox{}\hfill\begin{@IEEEauthorhalign}
}
\makeatother 

\IEEEoverridecommandlockouts
\usepackage{cite}
\usepackage{amsmath,amssymb,amsfonts}
\usepackage{algorithmic}
\usepackage{graphicx}
\usepackage{textcomp}
\usepackage{xcolor}
\usepackage{float}
\usepackage{booktabs}
\usepackage{array}
\usepackage{subfig}
\usepackage{siunitx}
\def\BibTeX{{\rm B\kern-.05em{\sc i\kern-.025em b}\kern-.08em
    T\kern-.1667em\lower.7ex\hbox{E}\kern-.125emX}}
\begin{document}

\title{Energy-Efficient Seizure Detection Suitable for low-power Applications\\

\thanks{This work has been partly funded by the German Federal
Ministry of Education and Research (BMBF) in the project MEDGE (16ME0530).}
}

\author{\IEEEauthorblockN{Julia Werner*}
\IEEEauthorblockA{\textit{Department of Computer Science} \\
\textit{University of Tübingen}\\
Tübingen, Germany \\
julia-helga.werner@uni-tuebingen.de}
*Corresponding author
~\\
\and
\IEEEauthorblockN{Bhavya Kohli}
\IEEEauthorblockA{\textit{Department of Electrical Engineering} \\
\textit{Indian Institute of Technology}\\
Bombay, India\\
bhavyakohli@iitb.ac.in}
~\\
\and
\IEEEauthorblockN{Paul Palomero Bernardo}
\IEEEauthorblockA{\textit{Department of Computer Science} \\
\textit{University of Tübingen}\\
Tübingen, Germany  \\
paul.palomero-bernardo@uni-tuebingen.de}
~\\
\and

\linebreakand

\IEEEauthorblockN{Christoph Gerum}
\IEEEauthorblockA{\textit{Department of Computer Science} \\
\textit{University of Tübingen}\\
Tübingen, Germany  \\
christoph.gerum@uni-tuebingen.de}
\and
\IEEEauthorblockN{Oliver Bringmann}
\IEEEauthorblockA{\textit{Department of Computer Science} \\
\textit{University of Tübingen}\\
Tübingen, Germany  \\
oliver.bringmann@uni-tuebingen.de}}

\maketitle

\begin{abstract}
Epilepsy is the most common, chronic, neurological disease worldwide and is typically accompanied by reoccurring seizures. Neuro implants can be used for effective treatment by suppressing an upcoming seizure upon detection. Due to the restricted size and limited battery lifetime of those medical devices, the employed approach also needs to be limited in size and have low energy requirements. We present an energy-efficient seizure detection approach involving a TC-ResNet and time-series analysis which is suitable for low-power edge devices. The presented approach allows for accurate seizure detection without preceding feature extraction while considering the stringent hardware requirements of neural implants. The approach is validated using the CHB-MIT Scalp EEG Database with a 32-bit floating point model and a hardware suitable 4-bit fixed point model. The presented method achieves an accuracy of \SI{95.28}{\percent}, a sensitivity of  \SI{92.34}{\percent} and an AUC score of $0.9384$ on this dataset with 4-bit fixed point representation. Furthermore, the power consumption of the model is measured with the low-power AI accelerator UltraTrail, which only requires \SI{495}{\nano\watt} on average. Due to this low-power consumption this classification approach is suitable for real-time seizure detection on low-power wearable devices such as neural implants.

\end{abstract}


\section{Introduction}
Approximately 65-70 million people are affected by epilepsy, one of the most common chronic neurological disease globally~\cite{moshe2015epilepsy}\cite{thijs2019epilepsy}\cite{thurman2011standards}. It is characterized by reoccurring seizures which result from an abnormal increased electric activity in the brain. They can lead to short-term mental absence, unconsciousness and convulsions\cite{engel2013seizures}\cite{fisher2005epileptic}. Reoccurring epileptic seizures not only diminish the overall quality of life of patients but are furthermore a severe safety hazard which can potentially cause severe accidents, in the worst case life-threatening.  One common treatment option are anti-epileptic drugs (AED); however, for approximately \SI{30}{\percent}~\cite{skarpaas2019brain} of adult patients these do not show any effect or are accompanied by non-bearable side-effects. 

For this group of individuals, other approaches need to be explored. One promising alternative is electric brain stimulation, which can be conducted through peripheral nerve stimulation, spinal cord stimulation or deep brain stimulation\cite{fisher2014electrical}~\cite{racine1972modification}. Devices for electric brain stimulation can either perform continuous stimulation or responsive stimulation upon seizure detection, and in both cases patients do not recognize the stimulation~\cite{bergey2015long}~\cite{fisher2014electrical}. For example, one such device for neuromodulation on the market is the Responsive Neurostimulation System (RNS) of Neuropace \cite{skarpaas2019brain} which is a neural implant with four electrodes that is implanted within the brain at a seizure onset region and recognizes as well as interrupts the beginning of a focal seizure. However, seizure devices are generally limited in size and therefore the available energy is limited. Each addition to the system requires its share of energy. If an algorithm is employed directly on the device, the total amount of area, number of computations and amount of energy has to be restricted as much as possible to limit the total energy consumption and prolong the battery lifetime.

This paper presents an energy-efficient approach to accurately determine anomalies within recorded Electroencephalography (EEG) data which are classified as seizures. By combining a light-weight neural network without preceding feature extraction but with subsequent time-series analysis methods, accurate anomaly/seizure detection is ensured while simultaneously providing a model with low complexity making it suitable for low-power applications. The model is validated using the CHB-MIT Scalp EEG Database  collected at the Children’s Hospital Boston\cite{shoeb2009application}\cite{physiobank2000physionet}. This enables subsequent deployment to low-power hardware accelerators for real-time seizure detection which is demonstrated on the UltraTrail hardware architecture \cite{bernardo2020ultratrail}.

\subsection{Related work}
Shoeb and Guttag published the CHB-MIT dataset in 2009 and achieved a sensitivity of \SI{96}{\percent} of $173$ test seizures with a mean latency of \SI{4.6}{\second}~\cite{shoeb2009application} while using Support Vector Machines (SVMs). Subsequently, others have used this dataset for further experiments but only few simulated the models on hardware. Hügle et al.\cite{hugle2018early} presented SeizureNet which was executed on hardware with a power consumption of \SI{850}{\micro\watt}.
Bahr et al.~\cite{bahr2021epileptic} tested the classification of a CNN and a low-power microprocessor with this dataset. They achieved a sensitivity of \SI{85}{\percent} with the CNN on a microcontroller and an average power of \SI{140}{\micro\watt}. Zanetti et al.~\cite{zanetti2020robust} used random forest models on an ARM cortex-M4 microcontroller and achieved a sensitivity of \SI{96.6}{\percent},  with a battery lifetime of 40.87 hours and \SI{7.34}{\milli\ampere}. Manzouri et al. tested a LSTM, CNN and a random forest classifier an ultra low-power microcontroller~\cite{manzouri2022comparison} with a dataset of the Epilepsy Center Freiburg resulting in \SI{7}{\micro\watt} for the CNN approach. 

For this application, some research has been conducted in combination with time-series analysis. For example, Craley et al.~\cite{craley2018novel} presented a coupled Hidden Markov Model for seizure detection and Dash and Kolekar~\cite{dash2020hidden} also performed seizure detection involving HMMs and Viterbi decoding, achieving an accuracy of \SI{96.87}{\percent}. However, this requires preceding feature extraction and furthermore, no hardware simulation was conducted and no energy estimation was provided in their experiments. Furthermore, there have been some usages of smoothing data with a moving average approach in the context of seizure detection in the past \cite{alotaiby2016channel} \cite{rana2012seizure} \cite{temko2011eeg} \cite{yu2020epileptic}. Rana et al.\cite{rana2012seizure} have employed the moving average method in the context of seizure detection. Although it was not used in combination with a CNN; but functioned as a tool for establishing a seizure detection threshold in their proposed method. Yu et al.~\cite{yu2020epileptic} combine a CNN for feature extraction with principal components analysis, Bayesian linear discriminant analysis and moving average filtering. Temko et al. \cite{temko2011eeg} use extracted feature-vectors with support-vector machines (SVMs) in combination with moving average smoothing.

We propose an energy-efficient and low-complex method which combines a very light-weight CNN with simple time-series analysis without the need of preceding feature extraction that is suitable for low-power hardware architectures. We validate this using the CHB-MIT dataset and the UltraTrail hardware accelerator.

\section{Methodology}\label{section:methods}

\begin{figure}[h!]
		\centerline{\includegraphics[width=0.45\textwidth]{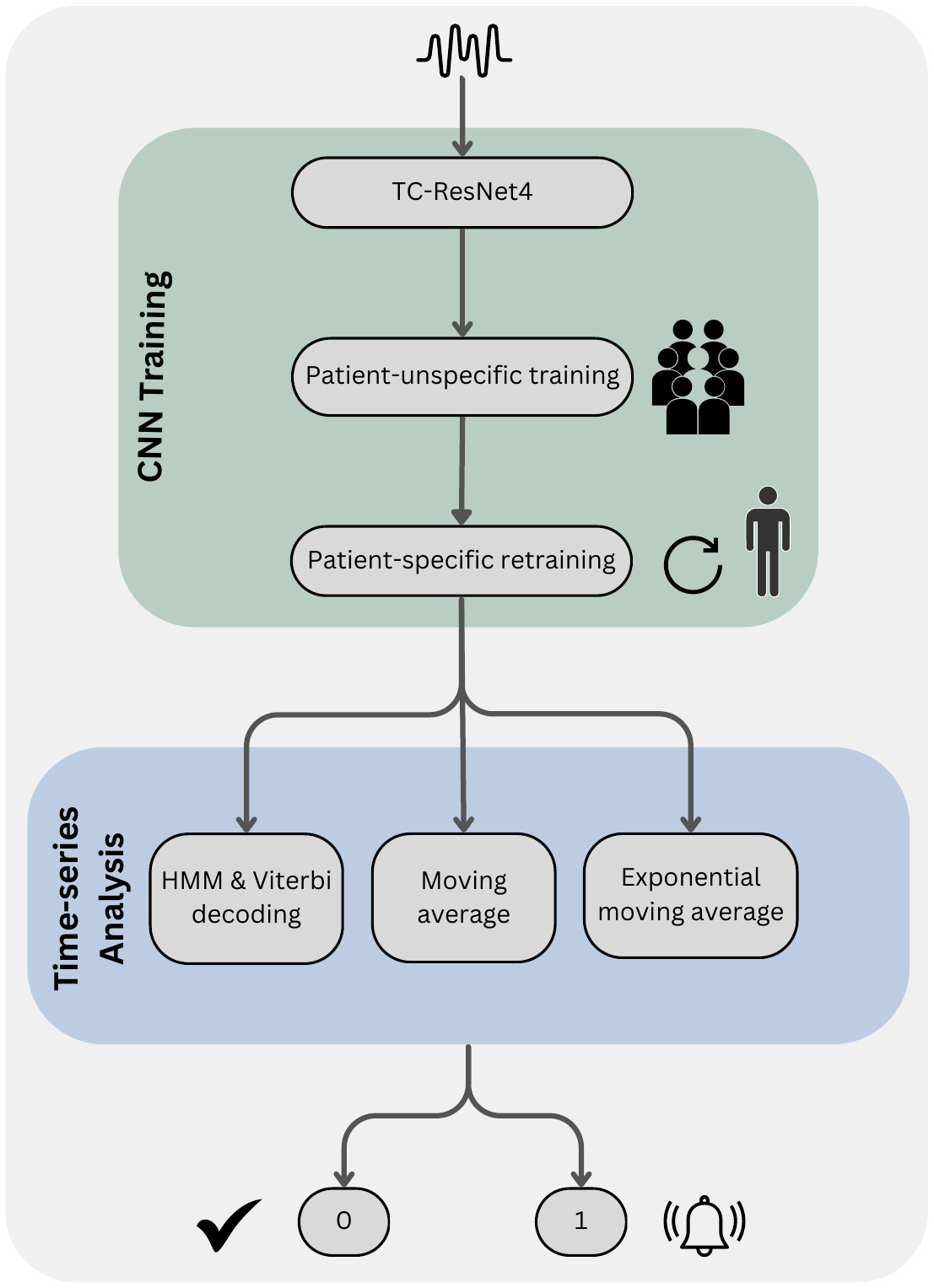}}
		\caption{Overview of the classification approach.}
		\label{fig:overview}
        \centering
\end{figure}

An overview of the complete classification approach is shown in Fig.~\ref{fig:overview}. First, the TC-ResNet4 is trained on preprocessed EEG data but without preceding feature extraction. After completion of the retraining, the model is used to classify incoming EEG data; subsequently, time-series analysis is employed either in the form of Viterbi decoding and a HMM or as a simple smoothing approach with simple moving average or exponentially moving average. Finally, for each EEG fragment, the classification approach returns either a $0$ for non-ictal recordings or $1$, if a seizure was detected.

\subsection*{Dataset description}

For all experiments, the CHB-MIT dataset\cite{physiobank2000physionet} \cite{shoeb2009application} was used, which contains the EEG data from 24 patients stored in the European Data Format (as \verb|.edf| files). In each patients' respective directory, there are multiple such \verb|.edf| files containing EEG data for up to 1 hour, read from different channels. The file annotations are provided separately, and the annotations contain the time-stamps for the start and end of a seizure.

\subsection{Preprocessing}\label{subsection:processing}

\begin{figure}[h!]
		\centerline{\includegraphics[width=0.5\textwidth]{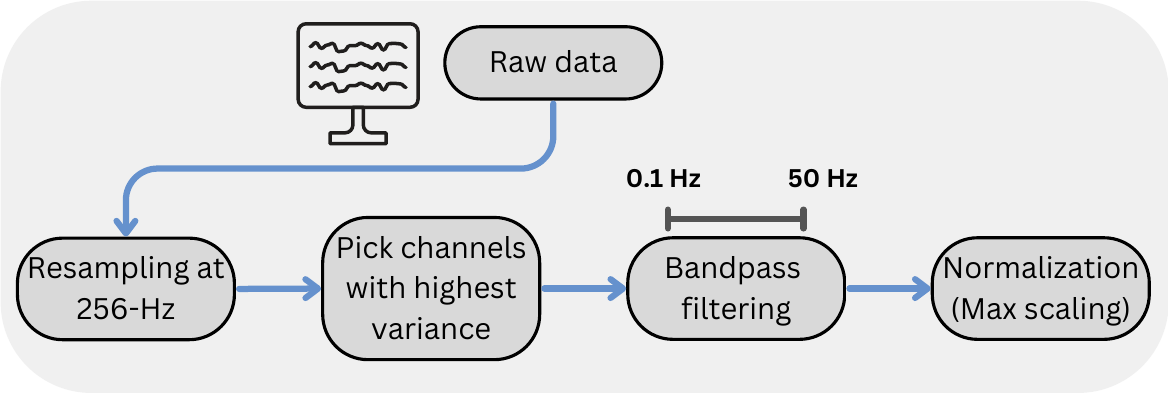}}
		\centering
		\caption{Preprocessing pipeline.}
		\label{fig:preprocessing}
\end{figure}  

To account for individual differences in EEG data, the main objective is to perform patient-specific retraining using a pre-trained base model and patient specific data. Thus, the data was split in the following way for each patient:

\begin{enumerate}
    \item The data was originally sampled at \SI{256}{\hertz}, any files with a different sampling rate were resampled to \SI{256}{\hertz}.
    \item Not all channels found in the dataset are present in each individual patient data. Therefore, only common channels across all patients were chosen to ensure uniform representation. From those, the number of channels was further reduced to $16$ based on the channels with the highest variance in the ictal data as suggested by~\cite{duun2012channel} to exploit hardware advantages.
    \item To remove DC components and the noisy component from the EEG measurement device, a band-pass filter was applied between \SI{0.1}{\hertz} and \SI{50}{\hertz}. Following the filter, the data was split into fragments of length \SI{0.5}{\second}, or correspondingly, fragments of $128$ samples. 
    \item The complete data for a patient was collected and labelled according to the extracted annotations, giving a large array of shape $(N\times128\times16)$. 
    \item Due to the scarcity of ictal data compared to non-ictal data, to maintain a reasonable ratio of positives and negatives, the non-ictal data was sampled such that for the dev and retrain set the final ratio of ictal to non-ictal data was $3$. However, this ratio was not enforced in the test set in order to test on a realistic proportion for each class.
    \item For the test set, for each patient one complete edf file consisting of an EEG recording with at least one seizure was collected.
    \item The remaining data was split into two stratified (to maintain the specified class ratio) subsets in a $40:60$ ratio. The two subsets are described below:
    \begin{enumerate}
        \item \textbf{dev (\SI{40}{\percent})}: Data in this subset is first collected from all patients, and then further split into a $80:20$ ratio (training and validation) for patient-unspecific training.
        \item \textbf{retrain (\SI{60}{\percent})}: Data in this subset is individually split into a $80:20$ ratio (training and validation), and is used for patient-specific retraining.
    \end{enumerate}
    \item The data in the \textbf{test} set is used purely for testing. The collected test data from all patients is used during the patient-unspecific training phase, and the individual patient data is used during patient-specific retraining. 
    \item For normalization, the absolute maximum value of the train set, validation set and test set was identified and the maximum value among those was used for normalizing each subset before training the neural network. 
\end{enumerate}

Fig. \ref{fig:preprocessing} illustrates the basic preprocessing pipeline for a given patient. The final class distribution for the complete train and test set, and average class distribution of retrain set and test set for all patients is reported in Table \ref{tab:class_distribution}. This also includes the data samples of the $2$ randomly selected patients, that is needed for training the time-series methods and thus excluded from final testing.

\begin{table}[htbp]
\caption{Class distribution of the (half second fragments) dataset used. \\{\small(* average data per patient)}}
\begin{center}
\begin{tabular}{lccc}
\toprule
\textbf{Subset} & \textbf{Negative samples} & \textbf{Positive Samples} & \textbf{Total Samples} \\ \midrule\midrule
dev           & 16,017                    & 5,339                    & 21,356                \\ 
retrain         &  24,111                    & 8,037                           & 32,148           \\ 
test            & 258,542                   & 4,796                     & 263,338  
        \\ 
retrain*     & 1,005                     & 335                     & 1,334               \\
test*        & 10,773                     & 200                       & 10,973                 \\ \bottomrule 
\end{tabular}
\end{center}
\label{tab:class_distribution}
\end{table}

\subsection{Classification approach}\label{subsection:approach}

\subsection*{Classification with a CNN}
Anomaly detection of EEG data is particularly useful in conjunction with a neural implant, so that it can act as a seizure detection device. However, the limited size of such implants restricts also the model size, e.g. in terms of the number of parameters. Thus, it is crucial to find a model which is suitable for low-power devices and simultaneously exhibits a sufficient accuracy and sensitivity. By combining temporal convolutional neural networks (TCNs) with residual networks (ResNets) the TC-ResNets\cite{choi2019temporal} are generated. TC-ResNets are light-weight 1D-convolutional neural networks with relatively few parameters which were successfully used for sensor-signal processing tasks\cite{choi2019temporal}. For this approach, it is beneficial that CNNs do not require a preceding feature extraction. Since TC-ResNets are characterized by a low complexity and additionally have been successfully applied to other problems~\cite{zhang2021autokws}, these CNNs were used as a base model. During the training process, it became apparent that a TC-ResNet4 provides the best trade-off between accuracy and complexity and was therefore used for further experiments without preceding feature extraction. The TC-ResNet4 archictecture \cite{choi2019temporal} is presented in Table~\ref{tc-resnet}.

    \begin{table}[htbp]
    \centering
    \caption{TC-ResNet4 architecture}\label{tc-resnet}
    \resizebox{0.48\textwidth}{!}
    {\begin{tabular}{lcll}
    \toprule
    \textbf{Layer Type}               & \textbf{Output} & \textbf{Params} & \textbf{MAC ops} \\ \midrule\midrule
    Input                                & $16\times128$ & 0                 & 0 \\
    Conv1d                            & $16\times 63$   & 768             & 48K              \\
    Conv1d                            & $24\times 32$   & 3,456           & 111K             \\
    BatchNorm1d                       & $24\times 32$   & 0               & 0                \\
    Hardtanh                          & $24\times 32$   & 0               & 0                \\
    Conv1d                            & $24\times 32$   & 5,184           & 166K             \\
    BatchNorm1d                       & $24\times 32$   & 0               & 0                \\ \midrule\midrule
    Conv1d                            & $24\times 32$   & 384             & 12.3K            \\
    BatchNorm1d                       & $24\times 32$   & 0               & 0                \\
    ReLU                              & $24\times 32$   & 0               & 0                \\ \midrule\midrule
    Hardtanh                          & $24\times 32$   & 0               & 0                \\
    TCResidualBlock                   & $24\times 32$   & 0               & 0                \\ \midrule\midrule
    GlobalAveragePooling1d & $24\times 1$    & 0               & 0                \\
    Dropout                           & $24$            & 0               & 0                \\
    Linear                            & $2$             & 48              & 48               \\ \bottomrule         
    \end{tabular}}
    \end{table}

The training was partitioned into two phases, considering at each step the suitability to implement this in hardware.

(1) The first phase of training involves using the collected data from all patients to train a patient-unspecific base model. Training was performed for $40$ epochs with a batch size of $128$ and the AdamW optimizer with a learning rate of $0.001$. For computing the loss, the different class frequencies were considered to calculate the weights for the loss function. Furthermore, we employed threshold moving~\cite{provost2008machine}~\cite{zhou2005training} to modify the default threshold of $0.5$ when classifying the data samples. To tune the threshold, we used the evaluations of the TC-ResNet4 on the train set by weighing the output probability of detecting a seizure with a weight $w \in \{1,2,3,4,5\}$. The smallest weight that achieved a sensitivity larger than $0.9$ in our experiments was $w=2$, which was therefore the hyperparameter of choice. Hence, this was employed for all following experiments, resulting in a slightly higher sensitivity at the cost of a marginally decreased specificity.

(2) For the patient-specific retraining in the second phase, the base model is loaded and then retrained on data from individual patients for additional $10$ epochs with the same configurations as the base model, except for a batch size of $8$. After the patient-specific TC-ResNet4 is fully trained, three different approaches of time-series analysis were used to enhance the given predictions of the TC-ResNet4. The experiments were conducted once with $32$-bit floating point computation as well as with $4$-bit fixed point representation to meet the hardware requirements. As a final step, time-series analysis is performed for each patient.

\subsection*{Classification with time-series analysis}
\subsubsection{Simple Moving Average}

The main reason to use subsequent time-series analysis was to improve the classification result of the CNN by smoothing the series of data points with a preferably computationally cheap method. One simple yet popular approach for this is applying a simple moving average (SMA) method~\cite{box2015time}. For this method, for a series of data points $X$, for a window size $w$ the mean is computed as follows:
\begin{align*}
    SMA &= \frac{1}{w} \sum_{n-w+1}^n X_i.
\end{align*}
In this setting, the SMA was applied with a window size $w=5$ to the final evaluations of the TC-ResNet4, the probabilities are finally converted to binary labels.\\

\subsubsection{Exponentially Moving Average}
The exponentially moving average\cite{hunter1986exponentially} (EWMA) provides exponentially weighted moving averages to put more emphasis on the latest observations of the neural network and was additionally applied. To determine the threshold for both methods, two randomly selected patients were used to obtain the best threshold based on a grid search and were further excluded from testing in all experiments.\\

\subsubsection{Hidden Markov Model}
Time-series analysis in the form of a Hidden Markov Model (HMM) and Viterbi decoding provides a hyperparameter free method, which therefore does not require many samples for training. This is a major advantage compared to other models such as LSTMs and the main reason for choosing this method. With this approach, the predicted labels of the TC-ResNet4 are passed on as integers to the HMM and subsequent Viterbi decoding returns the most likely sequence of hidden states based on the provided observations given by the TC-ResNet, similarly as employed in \cite{werner2023precise}. Due to the backtracking approach of the Viterbi algorithm, this detection method is accompanied by a delay which describes the time between the first occurrence of a seizure and the actual detection by the algorithm and mainly depends on the chosen window size for the included data. For the presented approach, the same window size $w=5$ was chosen as for the other methods, which results in a maximal delay of \SI{2.5}{\second} for \SI{0.5}{\second} fragments introduced by the Viterbi decoding. The transition and emission probabilities were acquired solely based on statistical measurements. To compute the transition probabilities, after the EEG data was split into \SI{0.5}{\second} fragments, the number of transitions from each class to the other class were counted for the data from the train and retrain set~\cite{kacprzyk2012advances}. The proportions can then directly be encoded in the transition probabilities of the HMM. The confusion matrix of the CNN classifying the training data in the last epoch of training naturally encodes the emission probabilities for this setting, thus the emission probabilities of the HMM were computed from this confusion matrix.

\subsection*{Metrics/Evaluation}
The accuracy, sensitivity, specificity as well as the false-positive rate (FPR) were computed based on the confusion matrix of the final model. Furthermore, the receiver operating characteristic curve area under the curve (AUC) score was computed with the scikit-learn metrics library.

\subsection{Inference}
To demonstrate the low-power consumption of this light-weight neural network for future applications, a low-power real-time AI accelerator for sensor-signal processing seems to be the most suitable hardware architecture. The AI accelerator UltraTrail, which previously has been successfully used for TC-ResNets on keyword spotting tasks \cite{bernardo2020ultratrail}, has been employed for the described neural network. One benefit of this hardware architecture is that it is optimized for TC-ResNet topologies, involving a configurable array of processing elemens and a distributed memory system with dynamic content re-allocation~\cite{bernardo2020ultratrail}.
The model was trained with quantization-aware training and finally, the PyTorch code was converted to C-code. For hardware execution, $4$-bit quantization was used for computing the weights, bias and features in fixed-point representation.

\section{Results}

As described in Section~\ref{subsection:approach}, the TC-ResNet4 was first regularly trained for $40$ epochs on data from all patients and then retrained on the data from each patient for $10$ epochs with a batch size of $8$ with \SI{0.5}{\second} fragments. First, the results for the non-quantized TC-ResNet4 with 32-bit floating point number computation is shown in Table~\ref{tab:results_non_quantized}. It is demonstrated that each of the three time-series methods provide proficient classification results.  The HMM provides the highest AUC score of $0.9530$, while the exponentially weighted moving average method provides the highest sensitivity with \SI{93.71}{\percent}. However, in terms of accuracy, specificity and AUC score, the exponential moving average method, which gives more weight to the latest observations, is characterized by a slightly poorer performance. One possible reason for this is that the window size of $5$ might be too restrictive for this method.
However, it is important to note, that although models are typically trained with $32$-bit floating-point number computation, which allows higher precision, this is not suitable for low-power hardware applications. To provide a hardware-suitable method, we explored how different word widths of the weights of the TC-ResNet affect the overall classification ability on these different approaches.

\begin{table}[htbp]
	\caption{Results of the non-quantized TC-ResNet combined with different subsequent time-series analysis methods.}\label{results}
    \centering
    \begin{tabular}{cccc}
        \toprule
         & \multicolumn{3}{c}{\textbf{TC-ResNet4 (32-bit floating-point)}} \\ \cmidrule{2-4}
         & Moving Average & Exp. Moving Average & HMM \\ \midrule\midrule
        Accuracy [\%] & 96.05 & 94.16 &  \textbf{97.84}\\
        Sensitivity [\%] & 92.04 & \textbf{93.71} & 92.67\\
        Specificity [\%] & 96.12  & 94.17 & \textbf{97.93}\\
        FPR &  0.0388 & 0.0583 & \textbf{0.0207}\\
        AUC score & 0.9408 &  0.9394 &  \textbf{0.9530}\\\hline
    \end{tabular}
\label{tab:results_non_quantized}
\end{table}

\begin{figure}[h!]
		\centerline{\includegraphics[width=0.45\textwidth]{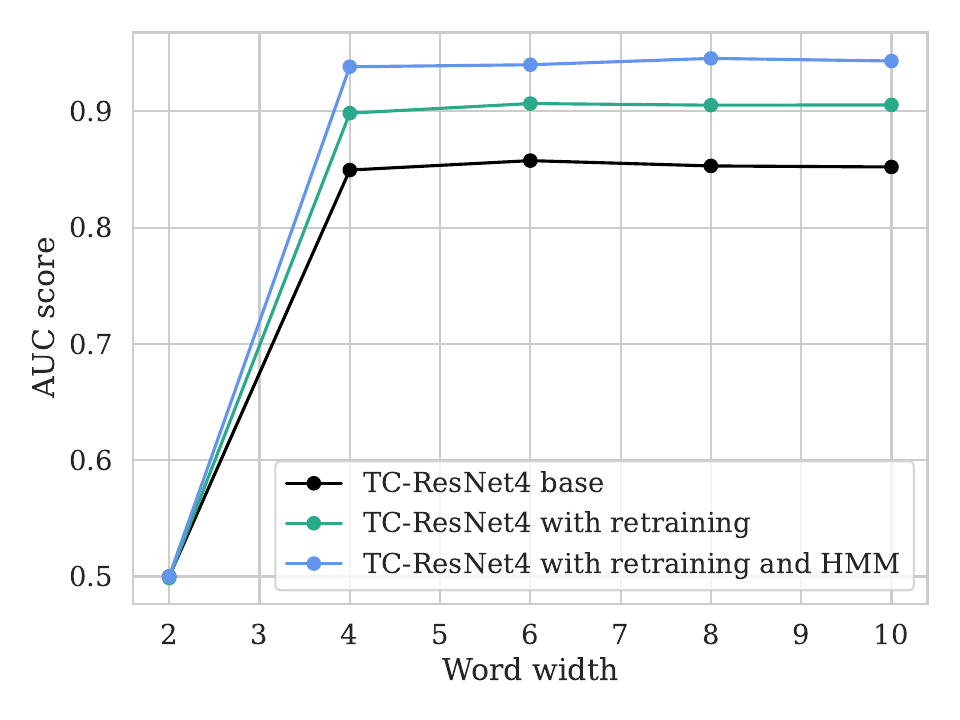}}
		\caption{AUC scores computed for the base model, the retrained model and the retrained model with subsequent Viterbi decoding while the weights of the model were quantized to different bits.}
		\label{fig:bitwidth}
        \centering
\end{figure}

The AUC score was computed for 1) the TC-ResNet4 base model (without retraining), 2) the same model with additional retraining, and 3) the retrained model with subsequent time-series analysis using Viterbi decoding after quantization-aware training was performed. The results are presented in Fig.~\ref{fig:bitwidth} for the word widths $\{2, 4,6,8,10\}$.
This demonstrates on the one hand that $4$ bits are sufficient to achieve good results and concurrently that the base model on its own is capable of good classification. Nevertheless, incorporating additional patient-specific retraining and adding time-series analysis increases the precision even further.

Subsequently, for the following experiments, quantization-aware training was performed for the base model as well as for the patient-specific model with a word width of $4$ bit allocated for each weight. Table~\ref{tab:results_quantized} presents the results of the trained and quantized TC-ResNet4 with the subsequent time-series analysis approaches as described in Section~\ref{section:methods}. All methods perform slightly weaker than if $32$ bits were used. This is expected due to a higher precision when computing with a larger number of bits. Nevertheless, incorporating quantization in the training process seems to compensate this and considering the hardware suitability, using $4$ instead of $32$ bits is a more reasonable approach for this setting.

    \begin{table}[htbp]
	\caption{Results of the quantized TC-ResNet4 combined with different subsequent time-series analysis methods.}\label{results}
    \centering
    \begin{tabular}{cccc}
        \toprule
         & \multicolumn{3}{c}{\textbf{ Quantized TC-ResNet4 (4-bit fixed-point)}} \\ \cmidrule{2-4}
         & Moving Average & Exp. Moving Average & HMM \\ \midrule\midrule
        Accuracy [\%] & 92.79 & 94.69& \textbf{95.28}\\
        Sensitivity [\%] & 92.15 & 90.20 & \textbf{92.34}\\
        Specificity [\%] & 92.80 & 94.77 & \textbf{95.34}\\
        FPR &  0.0720 & 0.0523 & \textbf{0.0466}\\
        AUC score & 0.9247 & 0.9248 & \textbf{0.9384}\\\hline
    \end{tabular}
\label{tab:results_quantized}
\end{table}

\begin{figure}[h!]
		\centerline{\includegraphics[width=0.45\textwidth]{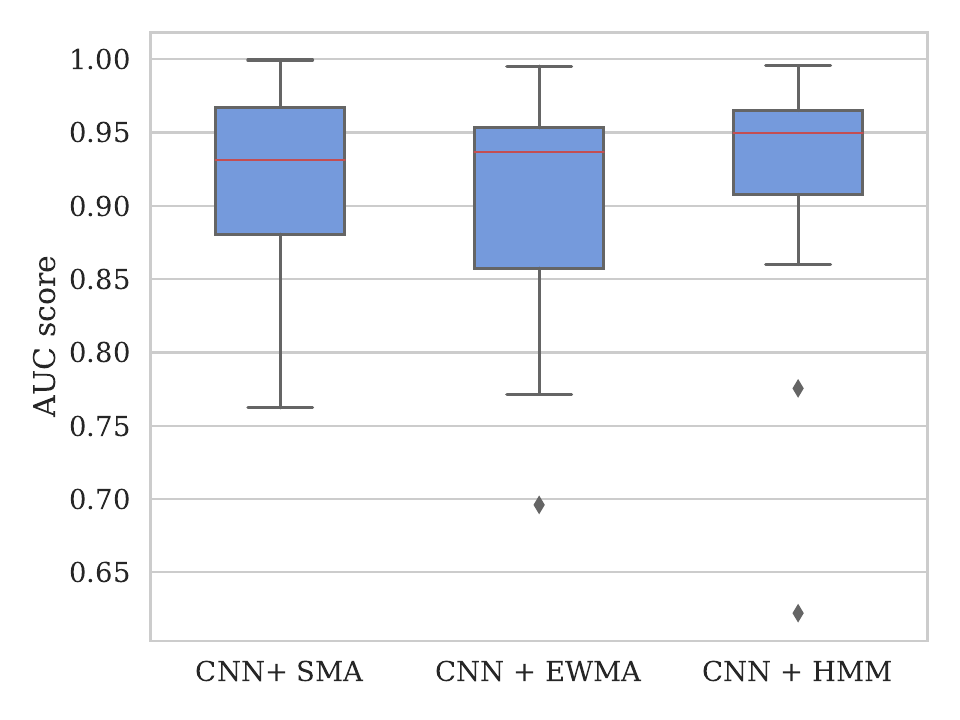}}
		\caption{AUC scores computed for each patient and for each of the three time-series analysis method: SMA, EWMA and HMM.}
		\label{fig:boxplot_auc}
        \centering
\end{figure}

For each metric, the HMM provides the best results compared to the moving average approaches with 4-bit fixed point representation. However, all three time-series analysis methods show similarly solid classification abilities. Additionally, for each patient the AUC scores were computed and plotted for each of the time-series methods as shown in Fig.~\ref{fig:boxplot_auc}. The main outlier was the patient with the ID 16, who also sticks out in the remaining results section.  

\begin{figure}[h!]
\centering
   \subfloat{\includegraphics[width=0.5\textwidth]{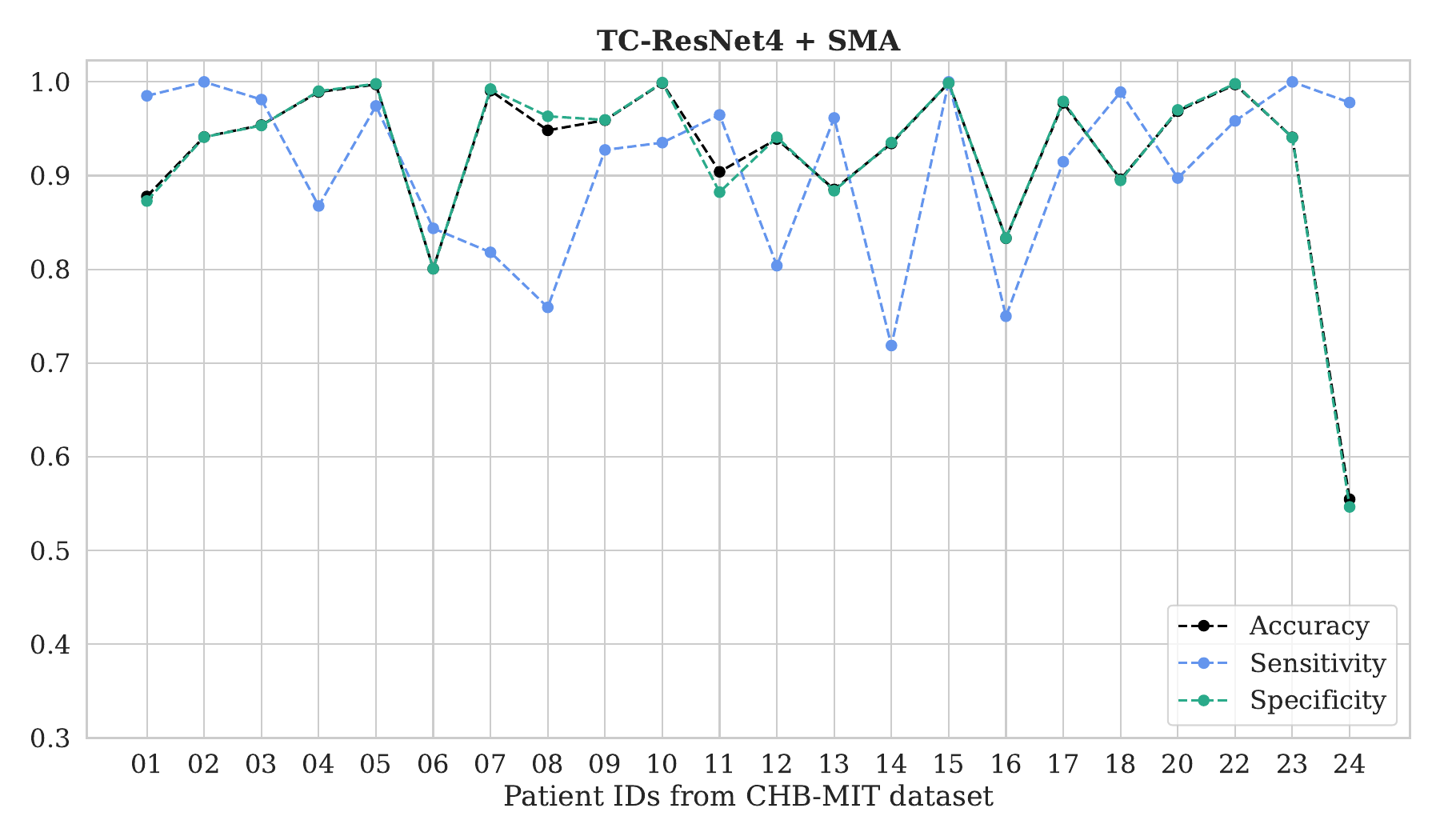}}\hfill
   \subfloat{\includegraphics[width=0.5\textwidth]{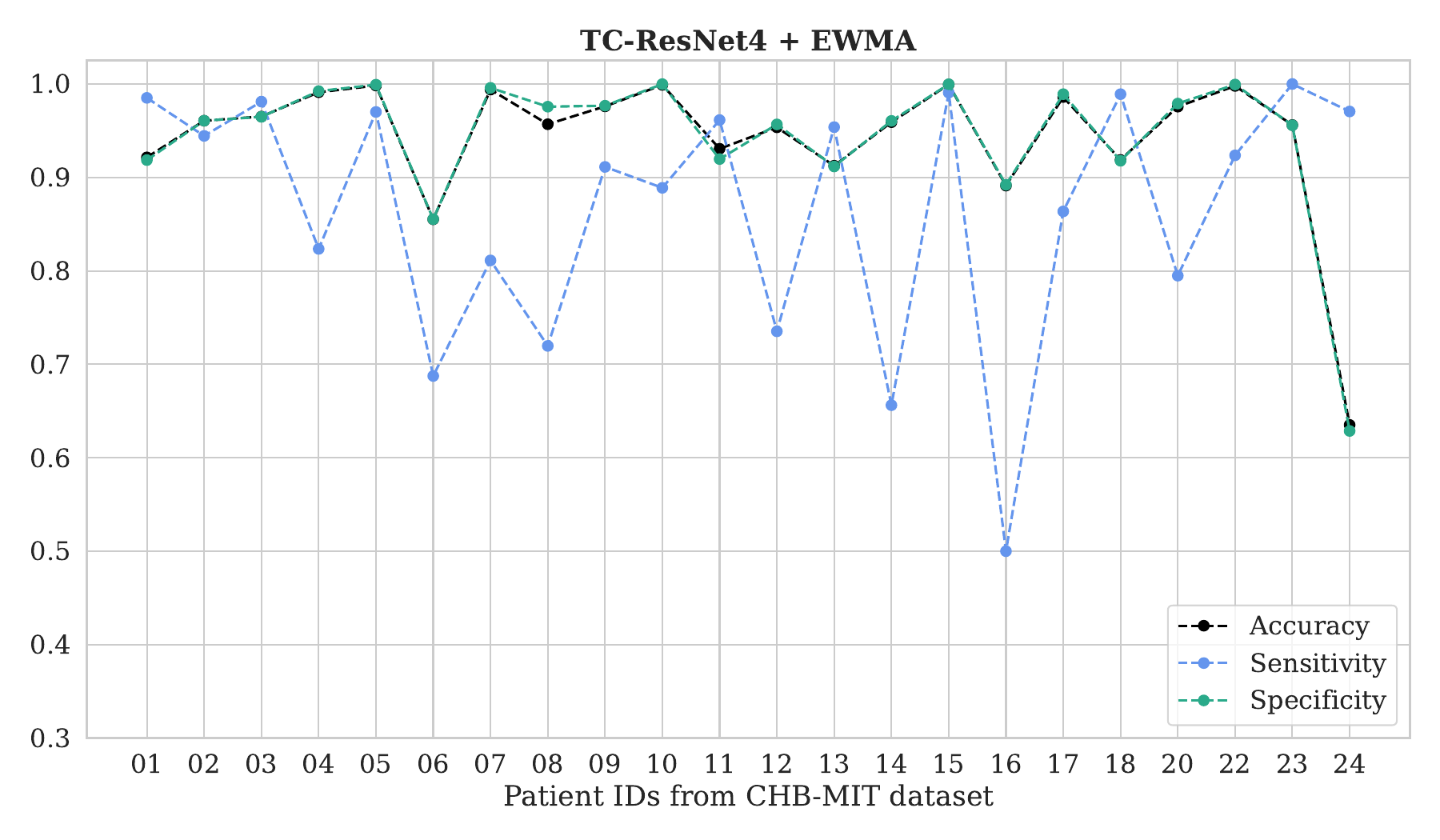}}\hfill
   \subfloat{\includegraphics[width=0.5\textwidth]{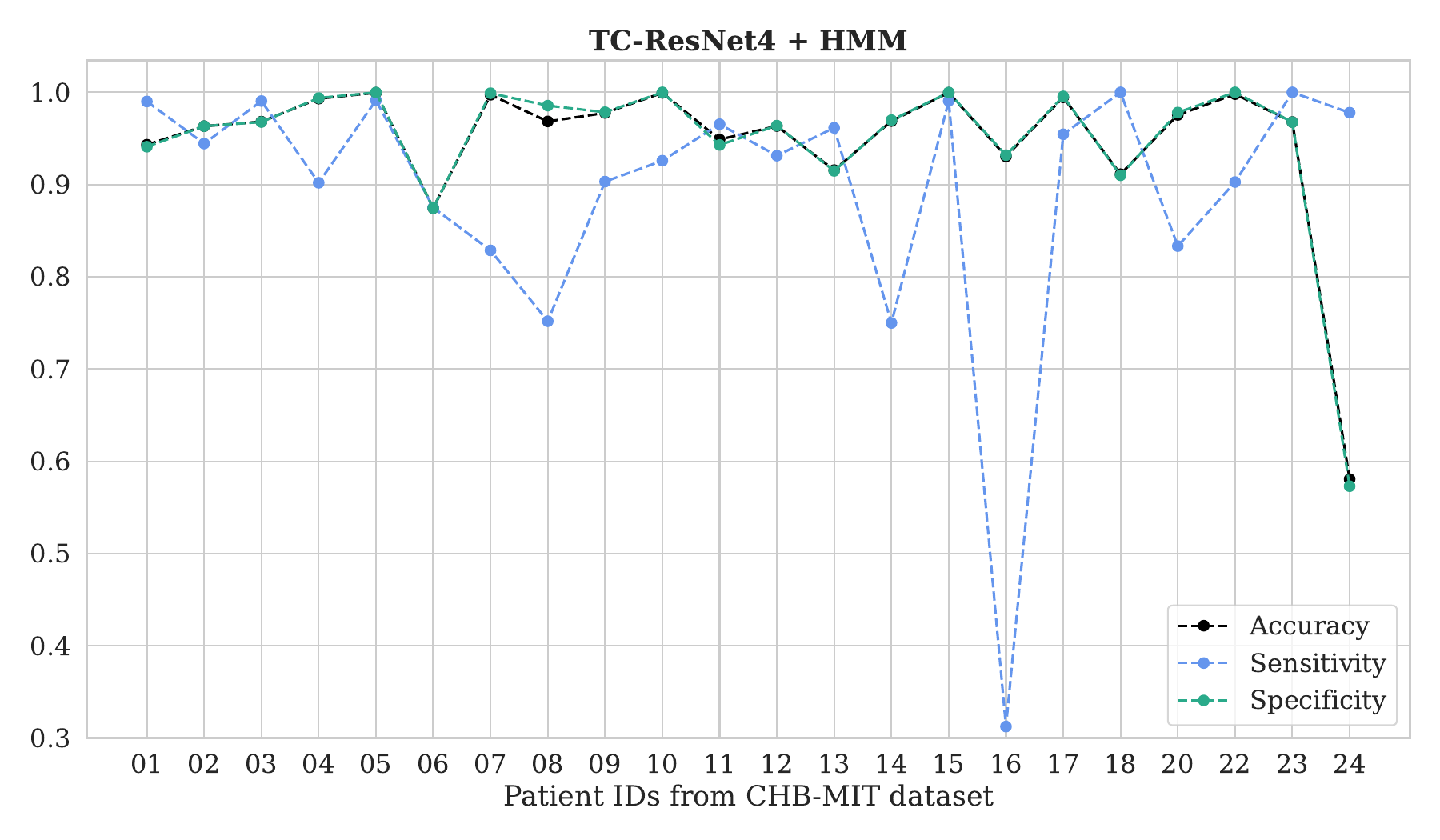}}
   \caption{Metrics shown for each patient and for each of the three approaches in combination with the 4-bit quantized TC-ResNet4.}
   \label{fig:metrics_plot}
\end{figure}

Furthermore, the sensitivity, specificity and accuracy were plotted for each patient for the classification approach involving the CNN and the different time-series methods as depicted in Fig.~\ref{fig:metrics_plot}. 
In particular, for the SMA and the EWMA, the sensitivity values are the lowest compared to the other metrics. Across all three time-series methods, the patient with the ID $16$ exhibits inferior results compared to the others. Interestingly, this patient also has the lowest number of seizure samples ($n=57$) in the retraining set, indicating that these are not sufficient samples to retrain the model perfectly. We further observe that the HMM in combination with Viterbi decoding performs slightly better across all patients (except for patient $16$) than the moving average methods. The mean detection delay of a seizure for the HMM approach was \SI{4.41}{\s}, including the delay introduced by the Viterbi decoding.

We validated that a patient-specific model achieves superior results compared to a patient-unspecific model. If this is realized in the future, it is important to note that the performance of the model improves with a larger amount of acquired data for each patient. The presented experiments suggest that it is beneficial to acquire some seizure data for each patient. On average, $335$ half-seconds of seizure data were used for retraining per patient in the presented experiments. However, if acquiring this data is not feasible, our results further indicate that simply using a trained base model in combination with the time-series analysis will provide sufficient results for the majority of patients as well.

\subsection{Inference}

The results of the Viterbi decoding for the described application can be directly implemented in a look up table (LUT) with a size of $k^{n+1}$, for $k$ states and a window size of $n$. There are $2^n$ possible combinations for the $n$ predictions of the CNN, which the HMM receives. For each sequence of predictions, a computed prediction of the HMM can be generated based on the Viterbi algorithm. The result for this can be stored as a $(n+1)^{\textit{th}}$ entry in the LUT. For our application we have two possible states (ictal, non-ictal) and a window size of $5$. Thus, the total size of this LUT would be $64$. 
The simple moving average method with a window size of $5$ requires the additional storage of $6$ integer values. Both methods only require a small amount of additional storage and basically no increased power consumption. Therefore, they are negligible and have not been included in the energy estimates.

\begin{figure}[h!]
		\centerline{\includegraphics[width=0.45\textwidth]{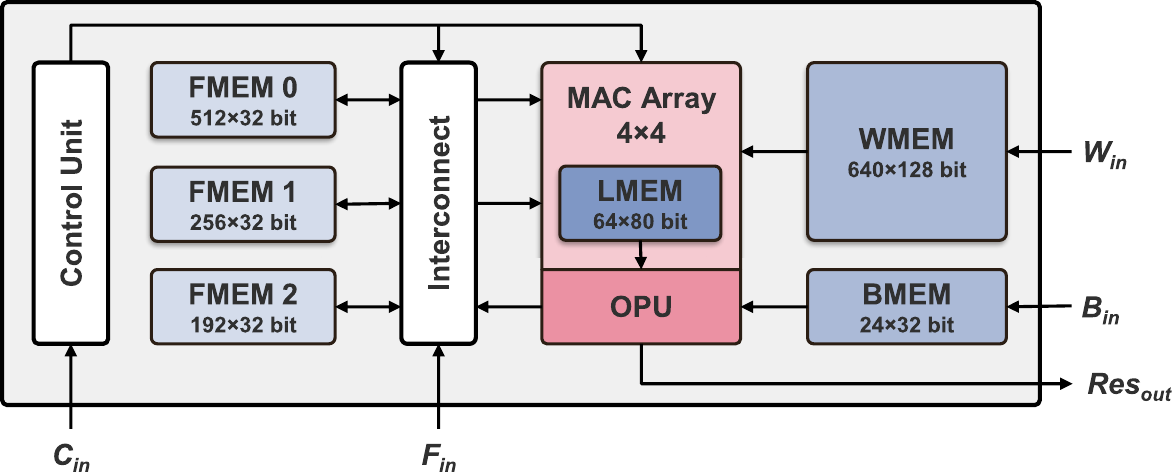}}
		\caption{UltraTrail architecture~\cite{bernardo2020ultratrail} with a $4\times 4$ MAC Array.}
		\label{fig:ultratrail}
        \centering
\end{figure}

The employed UltraTrail architecture is shown in Figure~\ref{fig:ultratrail}. This architecture consists of a control unit, three feature memories (FMEM), one interconnect unit, a local memory (LMEM), an output processing unit (OPU), the bias memory (BMEM) and a weight memory (WMEM) as published by Palomero et al.~\cite{bernardo2020ultratrail}. The FMEM stores the input and output features as well as the residual paths, while the WMEM and BMEM store all DNN parameters on chip. The multiply-and-accumulate units of the MAC Array compute the matrix multiplications of the network. The LMEM then stores the partial sums for the accumulation inside the MAC Array. The OPU is for example used for computing the activations, bias or pooling functions. In our experiments, the architecture is adapted to our specific use case and to the presented TC-ResNet4 by modifying the MAC Array to a $4 \times 4$ instead of a $8 \times 8$ MAC Array and adjusting the memory sizes accordingly, as shown in Figure~\ref{fig:ultratrail}.

\begin{figure}%
    \centering
    \subfloat{{\includegraphics[scale=0.24]{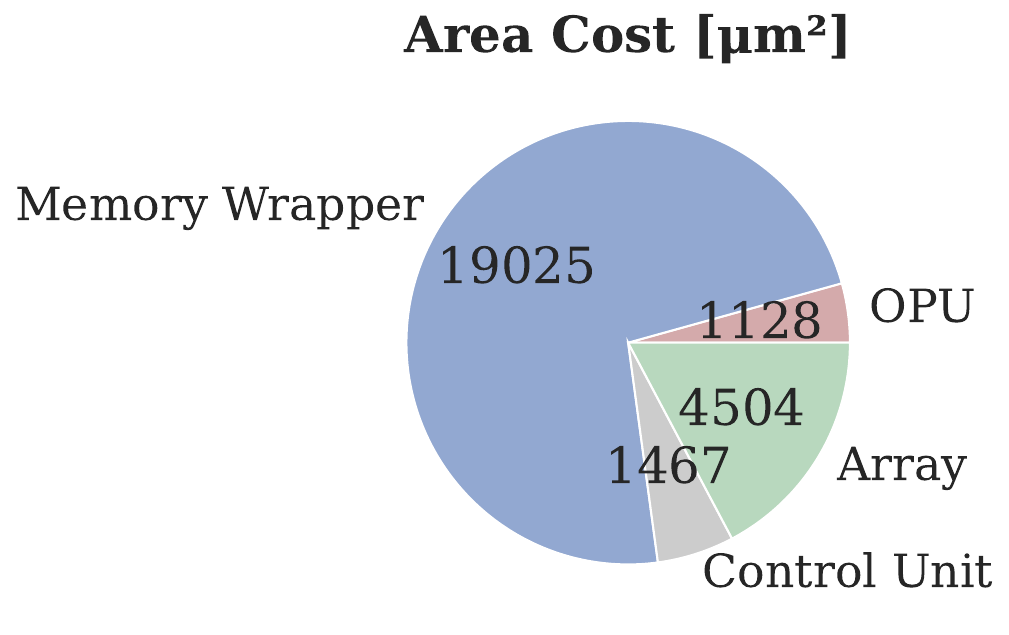} }}\hfill
    \subfloat{{\includegraphics[scale=0.24]{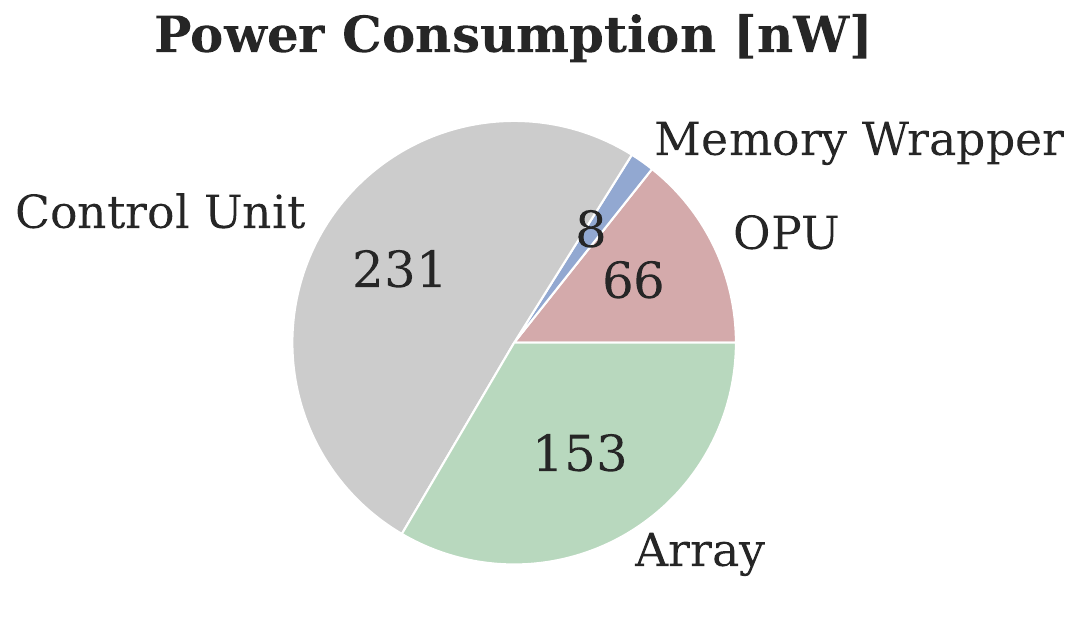} }}%
    \caption{Area costs and power consumption for different hardware components of the UltraTrail accelerator.}
    \label{fig:power-area}%
\end{figure}

For evaluation, the design was synthesized in GlobalFoundries 22FDX® \SI{22}{\nano\meter} FD-SOI technology using Cadence Genus. The power estimation was conducted for typical conditions (\SI{25}{\celsius}, \SI{0.8}{\volt}) using Cadence Joules.
Executing the quantized TC-ResNet4 on UltraTrail\cite{bernardo2020ultratrail} with $10$ inferences per second and a clock frequency of \SI{250}{\kilo\hertz} results in
a total power consumption of \SI{495}{\nano\watt} on average. The clock frequency can be adjusted depending on the real-time requirements of the use case. Executing the network requires \SI{80.626}{\milli\second}, afterwards the accelerator switches into a power gating state until the next inference starts. The total required area is \SI{26319.79}{\micro\meter^2}. Fig.~\ref{fig:power-area} depicts the area cost and the power consumption for each of the elements in detail. The memory wrapper consists of the FMEM, the WMEM, and the BMEM. In total, the CNN consists of 9840 parameters and requires $337 968$ MAC operations.\\

The measured power consumption in comparison to results from other work is listed in Table~\ref{tab:results}. However, comparing those power values is challenging, due to missing information or different types of measurement.
Hügle et al.~\cite{hugle2018early} used a MSP430FR599 microcontroller with a 32-bit hardware multiplier achieving \SI{850}{\micro\watt} for one inference. On UltraTrail, we need notably less power with only \SI{0.495}{\micro\watt} at \SI{10}{\hertz}, including the idle periods when the system is waiting for the next computation. 
Furthermore, our approach provides also advantages if used with other architectures as an alternative to UltraTrail. For example, our employed $4$-bit implementation is beneficial in combination with the Single Instruction Multiple Data (SIMD) technique. This theoretically allows the processing of $8\times 4$-bit operations in parallel on a $32$-bit hardware multiplier using a single $32$-bit register, which potentially leads to an energy reduction by a factor of $4$. 
Bahr et al.~\cite{bahr2021epileptic} use a CNN on a RISC-V based GAP microcontroller, which consumes an average power of \SI{140}{\micro\watt} for one inference on $1$ second EEG data. The execution time amounts to \SI{35}{\milli\second}. Although we have a higher execution time, our power consumption is notably less. Truong et al.~\cite{truong2018integer} compare 32-bit models to a 4-bit IntegerNet in their work, which demands $34-90$~\si{\micro\joule} for each inference. However, the specific hardware architecture used for measuring was not specified, which complicates a comparison. We have a total power consumption of \SI{0.495}{\micro\watt} with \SI{10}{\hertz} corresponding to an energy demand of \SI{49.5}{\nano\joule} for one inference including the idle time. Kiral-Kornek et al.~\cite{kiral2018epileptic} deployed their neural network to an ultra low-power TrueNorth chip and achieved a power consumption of \SI{<40}{\milli\watt}. Although they have not provided a concrete execution time, it is evident that our deployment combination is more efficient. Manzouri et al.~\cite{manzouri2022comparison} employed an ultra low-power Apollo 3 Blue microcontroller from Ambiq and had a power consumption of only \SI{0.495}{\micro\watt} for a random forest model and \SI{7.01}{\micro\watt} with a CNN. As they used a classification rate of \SI{1}{\hertz}, this corresponds to \SI{0.495}{\micro\joule} and \SI{7.01}{\micro\joule}, respectively. In their work it becomes evident that the random forest model performs inferior in the classification task as compared to the CNN. Nevertheless, our total energy demand including the idle time is notably less with only~\SI{0.0495}{\micro\joule}.

\begin{table}[htbp]
	\caption{Comparison of results to literature.}\label{results}
            \begin{center}
            \resizebox{\columnwidth}{!}{
			\begin{tabular}{lccc}
					\toprule
                    Authors & Model & HW-Architecture & Total power\\
                    \midrule\midrule
                    Hügle et al.\cite{hugle2018early} & SeizureNet &  MSP430FR599 & 850 $\mu$W\\
                    Bahr et. al\cite{bahr2021epileptic} & CNN & RISCV based GAP8 & 140 $\mu$W\\
                    Truong et. al\cite{truong2018integer} & IntegerNet & microcontroller & $34-90$ $\mu$J*\\
                     Kiral-Kornek et al.\cite{kiral2018epileptic} & DNN &  IBM TrueNorth &  $<40$ mW\\
                     Manzouri et al. \cite{manzouri2022comparison} & CNN & Ambiq Apollo 3 Blue & 7 $\mu$W\\  
					  This work & TC-ResNet4 & UltraTrail\cite{bernardo2020ultratrail} & \textbf{0.495 $\mu$W} \\
					\bottomrule
				\end{tabular}}
\end{center}
\label{tab:results}
*for each classification
\end{table}

\section{Conclusion}
Efficient seizure detection is essential for patients suffering from epilepsy who do not respond to drug treatment. In this work, a light-weight 1D-CNN with different time-series analysis techniques was combined, and validated using the CHB-MIT dataset achieving an accuracy of \SI{95.28}{\percent}, a sensitivity of \SI{92.34}{\percent} and an AUC score of $0.9384$ while allocating only $4$ bit per weight. The presented approach neglects the necessity of preceding feature extraction. Furthermore, it was demonstrated, that the classification model is suitable for real-time seizure detection by executing it on a low-power hardware architecture with a power consumption of only \SI{495}{\nano\watt} on average for one inference including the remaining idle time with \SI{10}{\hertz}. Especially compared to current baselines for this classification task, the low-power consumption with this approach stands out.

\bibliographystyle{splncs04}
\bibliography{thebibliography}

\end{document}